\documentclass[11pt,a4paper]{article}
\usepackage[utf8]{inputenc}
\usepackage[pdftex]{graphicx}
\pdfoutput=1

\def\affiliation#1{\gdef\@affiliation{#1}}
\def\@affiliation{\@latex@error{No \noexpand\affiliation given}\@ehc}

\title{Modelling the Evolution of Spatially Distributed Populations in the Uniformly Changing Environment - Sympatric Speciation}

\author{Wojciech Waga, Marta Zawierta, Stanis\l{}aw Cebrat}

\affiliation{Department of Genomics, Faculty of Biotechnology, University of Wroc{\l}aw, ul. Przybyszewskiego 63/77, 51-148 Wroc{\l}aw, Poland. E-mail: cebrat@smorfland.uni.wroc.pl.}

\begin{document}

\maketitle

\begin{abstract}We have simulated the evolution of age structured populations whose individuals represented by their diploid genomes were distributed on a square lattice. The environmental conditions on the whole territory changed simultaneously in the same way by switching on or off some requirements. Mutations accumulated in the genes dispensable during a given period of time were neutral, but they could cause a genetic death of individuals if the environment required their functions again. Populations survived due to retaining some surplus of genetic information in the individual genomes. The changes of the environment caused the fluctuations of the population size. Since the simulations were performed with individuals spatially distributed on the lattice and the maximal distance between mating partners was set as a parameter of the model, the inbreeding coefficient in populations changed unevenly, following the fluctuation of population size and enhancing the speciation phenomena. 
\end{abstract}

\section*{Key words:} sympatric speciation, age structured populations, genetic complementation, evolution, Monte Carlo simulations.


\section{Introduction}

A species is defined as a group of organisms interbreeding in natural conditions and producing fertile offspring while a speciation is a process of species formation. There are two distinct ways of a speciation; allopatric speciation when the population of an original species is divided into two or more groups separated by physical, geographical, environmental or biological barriers and, sympatric speciation when a new species emerges inside the original species population without any former barriers. It is easy to imagine the allopatric speciation which could be the result of the accumulation of differences in the genetic pools of separated sub-populations. Though, it was difficult to find the evolutionary mechanisms leading to the sympatric speciation basing on the mean field models of Mendelian populations used in the neo-Darwinian theory of evolution. The homogenization of the genetic pool of Mendelian population should effectively prevent any sympatric speciation. That is why sympatric speciation has been considered for a long time as the process which could be neglected in a building up the biodiversity \cite{Mayr}. Nevertheless, the models describing the processes of the genetic pool evolution of the Mendelian population have assumed some simplifications which formerly seemed to be justified and, in fact, might be used in some instances. When analysing the phenomenon of sympatric speciation, a few of such assumptions might not be introduced into the models. 

It is assumed that Mendelian population is large (infinite) and panmictic - each individual in the population may look for a sexual partner in the whole population and there are no mating preferences. In such populations the Hardy-Weinberg principle (HWP) is in force \cite{Hardy}. HWP states that allele frequencies in a population remain constant or are in equilibrium from generation to generation. HWP can be used, for example, for calculation of the frequency of defective recessive alleles based on the frequency of autosomal diseases. The predictions of the HWP are affected if the analysed allele or haplotype are under a selection pressure, a mutational pressure, there is a non-random mating, or the population size is limited. Furthermore, the Fisher-Write (F-W) model introduced another simplification based on the Mendelian principle saying that alleles can assort to the gametes independently \cite{Fisher}. In nature, the large groups of genes are located on single chromosomes. Since the crossover frequency is relatively low, such groups are often transmitted to the gametes as clusters and genes in the cluster definitely are not transferred independently. There are some data suggesting that even genes located on different chromosomes are not transferred into gametes independently \cite{Ehlers}, \cite{Trowsdale}, \cite{Younger}, \cite{Hiby}, \cite{Gendzekhadze}, \cite{Yawata} or there are expected some mechanisms of the gamete recognition which could be a further restriction imposed on the independent assortment of genes \cite{Stauffer}. The analogous mechanism is well known in bacteria as an entry exclusion system \cite{Novick}. If we assume that individuals, even in large populations, are spatially distributed, then, it is natural to impose some restriction on distances where they can look for mating partners and where they can place their newborns. In such a locally limited mating, the effective size of a population is not equal to the size of the whole panmictic population even if there are no other mating preferences \cite{Waga}. In such populations, there is a very strong relation between the effective size of population (which defines the inbreeding) and the intra-genomic recombination rate \cite{Marta}, \cite{Zawierta}. The probability that a newborn would receive a long non-disrupted string of genes or even two homologous fragments of haplotypes from one ancestor depends on both the intra-genomic recombination rate and the effective size of population \cite{Marta}. If such parameters are considered in the models of a population evolution, the sympatric speciation appears as a very common phenomenon. Furthermore, it seems to be an inherent property of expanding populations, especially at the borders of such populations \cite{Waga}. 

The other problem is a response of organisms to the environmental changes. In the post-genomic era, many experiments suggested that genomes include dispensable genes. For dozens of percent of genes, it is very difficult to find any phenotypic trait in the individuals with one of those genes eliminated from their genome or silenced. Though, for a fraction of such individuals with one knocked out gene, a specific phenotype can be found if many alternative environmental conditions are checked \cite{Giaever}, \cite{Kamath}, \cite{Kobayashi}. These experiments indicate that some genes in the genomes may not be used during the whole life span of an organism if the environmental conditions have not required their specific function. If organisms have lived in such conditions for many generations, the accumulation of mutations and eventually the deletion of the gene and its function could be expected for a substantial fraction of population. If the environment turns to demand the function again, this fraction of population could be in risk of genetic death. Computer models have shown that populations considerably shrinks in such conditions passing through the so called bottle neck or could be even extinct \cite{Fabian}.

In this paper we describe the results of simulations suggesting that environmental changes inducing the fluctuations in size of populations could enhance the probability of the sympatric speciation. In extreme, a catastrophe, considerably reducing the population size and increasing the inbreeding, could be a natural cause of a very fast speciation, like that one observed in Permian. 

\section{Materials and Methods}
We have used the diploid version of the Penna model \cite{Penna} for simulation of the age structured populations. The Penna model, based on the Monte Carlo method, has been used in more than a hundred works studying different aspects of the biological evolution (for review see \cite{Stauffer1}). In our modification, a population was composed of individuals represented by their diploid genomes composed of two pairs of bit-strings (chromosomes) - one 128 bits long, the other one 384 bits long. Bits correspond to alleles. If bit is set to 0 it corresponds to the wild (correct, functional) allele if it is set to 1 it corresponds to the defective allele. All defective alleles are recessive - both alleles at the corresponding loci have to be defective to determine a defective phenotypic trait. Declared number T of defective phenotypic characters kills the individual. Bits are switched on chronologically - the consecutive bits are switched on in the consecutive Monte Carlo steps (MCs) of the individual’s life; single bits in the same positions (loci) of two short bit-strings and three bits in the corresponding positions of each of two longer bit-strings are switched on in one MCs. Thus, older individuals have more bits switched on. The Penna diploid model reproduces very well age structures of natural populations including the human populations and their changes during the last century \cite{Bonkowska}. 

Individuals were placed on a square lattice of size a x b, no more than one individual per square. A female in the reproduction age looks for male partner being also at the reproduction age at the distance no larger than P. To produce the offspring, both parents copy their genomes. During the genome replication one mutation is introduced into each haplotype at the random position (the frequency of mutation per bit is the same for both bit-strings). Mutation changes bit 0 to 1, if the chosen bit is already 1 it stays 1 (there are no reversions). Homologous bit-strings recombine at the random position with probability C in the process mimicking crossover. Notice, a recombination rate for both pairs of bit-strings is the same. Single copies of both bit-strings (or the products of their recombination) form a gamete. Gametes of two partners fuse forming the genome of a zygote. The zygote is going to be born if it does not express T phenotypic traits in the first 50 loci of the short bit-strings and 150 loci in the longer ones, otherwise it dies (this death corresponds to miscarriage). The other condition for delivering a newborn is a free square at the distance no larger than B from a mother. If there is no free square at such a distance a baby is not born. The alleles placed on the consecutive loci (the single locus of the shorter chromosome pair and the three loci of the longer pair) of the newborn genome successfully placed on the lattice are switched on and checked every MCs until it reaches the reproduction age R (80). During the reproduction period each female can deliver up to b newborns at each MCs. The maximum life span of individuals is 128 MCs, when all bits have been switched on. Under parameters used in the simulations, individuals die earlier because of their genetic conditions. In the Penna model, a characteristic gradient of defective genes is generated - the higher fraction of defects occurs in genes expressed later after a minimum reproduction age, for the last loci in the bit-strings the fraction of defective alleles reaches 1.The visualization of the population evolution on the lattice has been done by colouring the squares occupied by individuals according to their genotypes. To each number $1-2^{24}$ a specific colour has been ascribed. The central parts of the two longer bit-strings of an individual (bits 105 - 128) have been read and each fragment of the bit-string was considered as a number. The squares occupied by the individual were coloured according to the larger of the numbers. 

To introduce environmental changes, a pattern of the environment is defined in the bit-strings 512 bits long, corresponding to all loci of both chromosomes. Environmental changes were introduced by switching the requirement for a function - on (value of bit 0), or off (value of bit 1). If the requirement for function is switched off, the activity of the corresponding locus on a chromosome has a neutral character and even if both bits at the position are set to 1 the defective phenotypic trait of the position is not seen - an individual survives. Those positions can accumulate defects without any penalty by selection. Nevertheless, after a long period of neutrality of a given position, a substantial fraction of population could loss the corresponding function and switching on the requirement for this function could be deleterious for individuals which have already lost it. Environment has been changed in two different ways: 

\begin{itemize}

\item {systematically - bit-strings describing the environment condition have been divided for fragments 24 bits long. At each of the fragment the first 8 bits were set to 1 for a declared period of time. Then, those bits were set to 0 again while the consecutive 8 bits are set to 1 and, in the third period the second group was set to 0 while the last 8 bits were set to 1. Such a round of three changes was periodically repeated. In this version, the frequency of environmental change and the position of neutral bits have been precisely controlled.}

\item{randomly - declared number of 512 bits of the environment pattern is changed at each MCs from 1 to 0 or from 0 to 1. Notice that after simulations long enough half of bits are 1 and half are set to 0 which means that half of genetic information in the genomes is dispensable at a given MCs.}
\end{itemize}

\section{Results and Discussion}
The monte Carlo models of population evolution in the changing environments were used several times \cite{Brigatti}, \cite{Sa Martins}. Nevertheless, consideration of the genome structure of individuals of evolving populations generates some problems. Usually a pattern of the environment is constructed in the same way as the genomes or the haplotypes of individuals - as the bit-strings. Bits representing genes in the genome can mutate from 0 (functioning) to 1 (defective). Reversions in nature are very rare and in the models it is often assumed that there are no reversions. There is a quite different situation in the description of the environment conditions. The environmental changes are set in the models by replacing the values of bits in the environmental pattern ($0\Leftrightarrow1$) \cite{Sa Martins}. It is natural that the environmental conditions can fluctuate and ,,reversions'' should be considered, usually with the same frequency as in the ,,forward'' direction. If a bit in the environment pattern is set to 0, it means that the required status of the corresponding, functional allele is also 0. Such a form of the functional allele is usually considered as a dominant one; it can complement the defective allele in the homologous locus of the second haplotype. If a bit 0 is replaced by 1 in the pattern, it means that environment requires 1 in the genome in the corresponding locus. That raises a problem - how to set the character of bits 0 - should they stay dominant or switch to recessive in the changed environment? If 0 is still dominant, it means that the mutation of the allele 0 which now is ,,a bad one'' to the ,,good allele'' (1) is recessive and a mutation from the good allele to the bad one is dominant. In fact, mutations in the most of loci are recessive. Recessive mutations lead to a loss of allele function, but usually they are masked if a wild copy of the allele is present in the genome. If the probability of changing the bits in the environmental pattern in both directions is equal - the number of loci where mutations are dominant or recessive in equilibrium would be equal, too. This condition itself changes the characteristic of the genetic pool and the population evolution considerably because it leads to a situation where deleterious mutations are dominant in 50\% of the loci. The other problem is in the way, how genomes in the evolving populations follow the changes in the environment patterns. The 0 bit in the pattern requires 0 in the genome, the change to 1 in the pattern changes the demand to the form 1 of the functional allele. The reversion of the bit in the environmental pattern demands the reversion of the allele in the genome (by mutation). This property of the model is unrealistic because it assumes that mutations change the gene from one functional form to the other one, also functional and even more unrealistic - next mutation changes it to the former one. In fact, in such a model, there is a switching between two conditionally functioning forms of genes just by single mutations. 

To omit all these obstacles, we have assumed that environmental conditions are described only by the requirement of a given function or not. If the function is dispensable - the genetic activity of the corresponding locus has the neutral effect. On the other hand, mutations always transform the functional gene into its defective, non-functional form, there are no reversions and the functional allele can complement the defective one. 

\subsection{Constant Environment}
\begin{figure}
\includegraphics[angle=0,width=\textwidth]{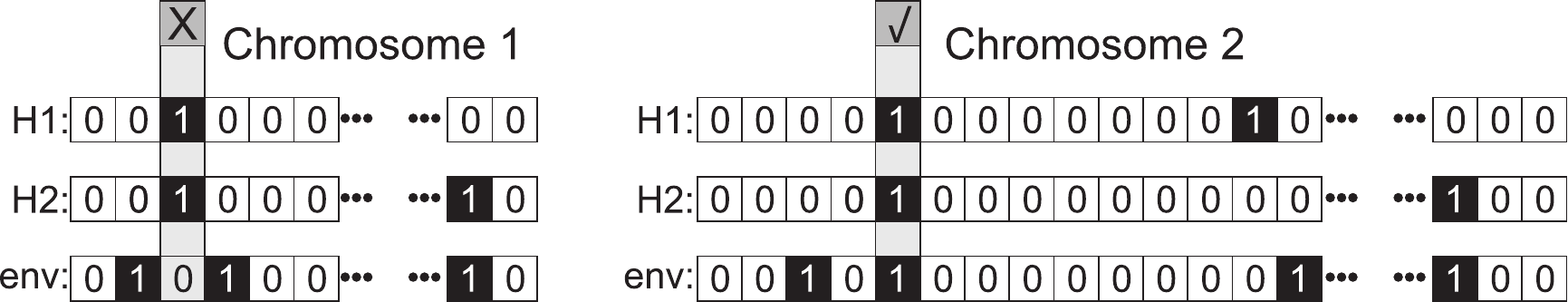}
\caption{The scheme of the relation between the genomes and the environment. Individuals are represented by a pair of chromosomes 1 - bit-strings 128 bits long, and a pair of chromosomes 2 - bit-strings 384 bits long. One bit of chromosome 1 and three bits of chromosome 2 are switched on at each Monte Carlo step. The environmental pattern is represented by a bit-string 512 bits long (lower part), each bit corresponding to one bit in chromosomes. The selection kills the individual only if both alleles in the same locus are defective (set to 1) and the environment is set to 0. Even if both alleles are set to 1, an individual is not killed if the environment does not require the function (1 in the environment pattern).}
\label{f1}
\end{figure}
\begin{figure}
\includegraphics[angle=0,width=\textwidth]{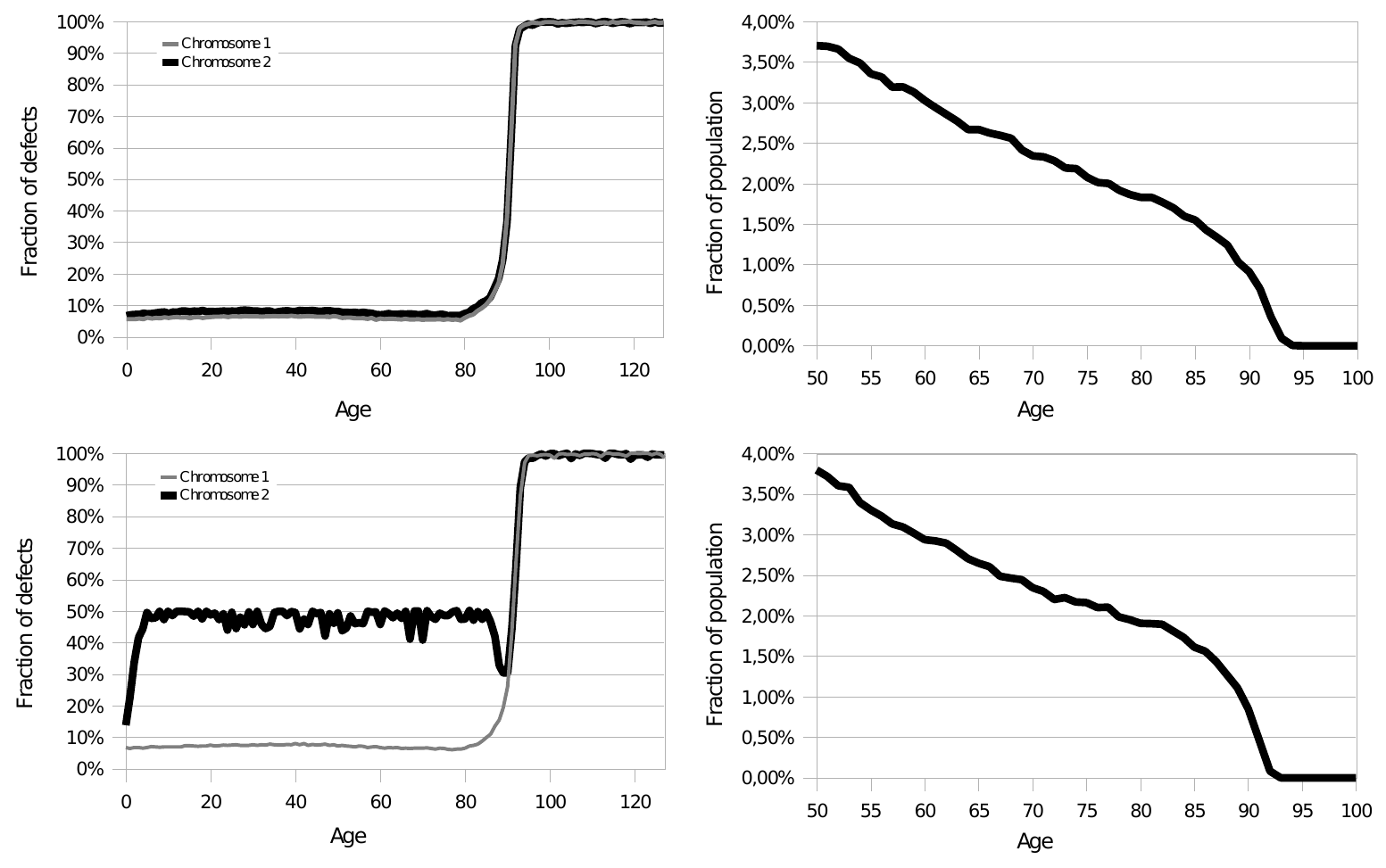}
\caption{The distribution of defective genes on chromosomes (on the left) and age distribution of simulated populations (on the right) after simulations under different recombination rate; upper panel C=1.0 and lower panel C=0.34. For higher cross over rate the fraction of defective genes and their distribution along the chromosomes is almost the same. For lower recombination rate shorter chromosome has lower fraction of defective genes while the longer one is already in complementation strategy and has higher fraction of defects unevenly distributed. Notice that three bits of longer chromosome are switched on at each MCs, the age distribution is shown only for individuals at the age 50 and above (for more information please refer to text).}
\label{f2}
\end{figure}
\begin{figure}
\includegraphics[angle=0,width=\textwidth]{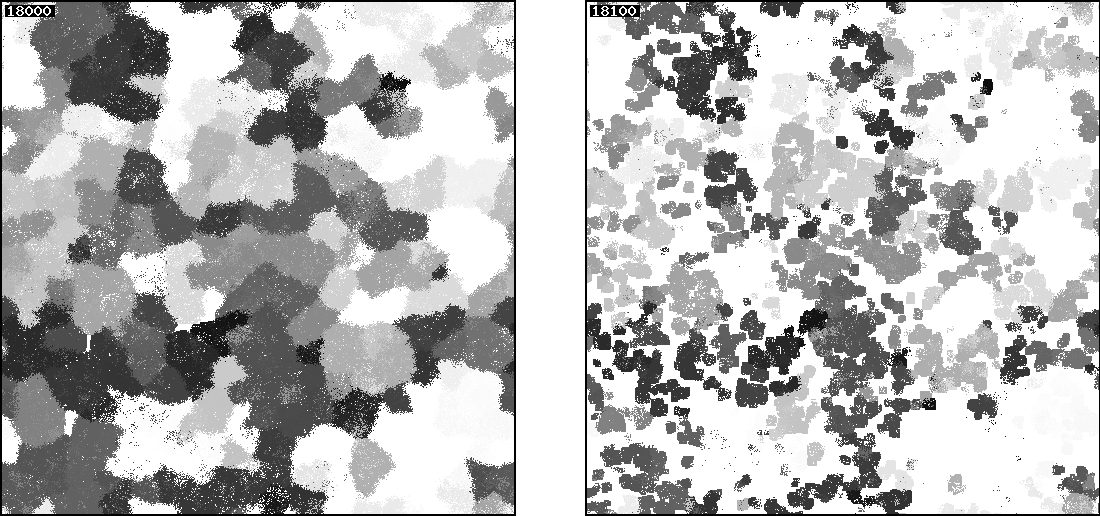}
\caption{Distribution of population on lattice close to the critical values of parameters. Left panel just before the environmental change and the right one 100MCs (one generation) after the change.}
\label{f3}
\end{figure}
In the Fig. \ref{f1} the schemes of the chromosomes and the environmental pattern are presented. Our control simulations start with genomes loaded with 7,5\% of randomly distributed defective genes and the environment demanding the function of all loci.  If the recombination rate is relatively high (C=1), the population is highly polymorphic and 98\% of available territory is occupied. The age structure of this population and the average fraction of defective alleles along the chromosomes are shown in Fig. \ref{f2}. Under the lower recombination rate, the longer chromosomes choose the complementing strategy while the shorter bit-strings stay in the purifying selection (the low fraction of defective alleles). The complementing fragments of haplotypes have the unique distribution of defective alleles and force the speciation. In the snapshot of the population evolution the speciation is illustrated by large spots of the same colours (Fig. \ref{f3}). The individuals with the same colours have the same configuration of wild and defective alleles in the central parts of their genomes (see the model section). There is no specific distribution of alleles located on the shorter bit-string. The same population coloured according to the central part of the short bit-strings is polymorphic (not shown). It could be interpreted that genes located on the shorter bit-strings, with relatively low number of alleles per genetic unit (centiMorgan) are not involved in the sympatric speciation. 

\subsection{Changing Environment}
\begin{figure}
\includegraphics[angle=0,height=8cm]{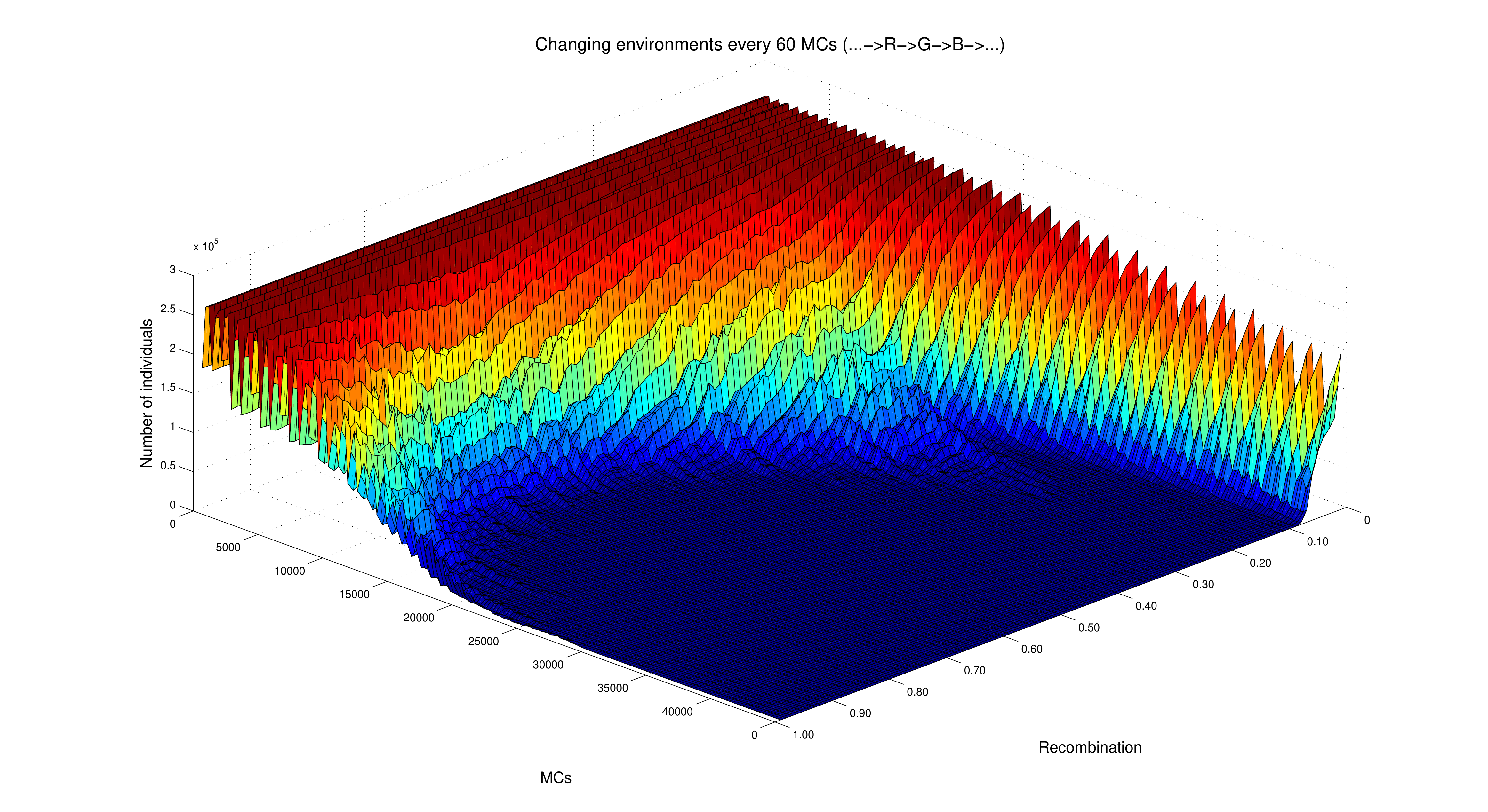}
\caption{Fluctuation of the population size under cyclic changes of the environment - one third of functions neutral for 60 MCs (see text for more details).}
\label{f4}
\end{figure}
\begin{figure}
\includegraphics[angle=0,width=0.5\textwidth]{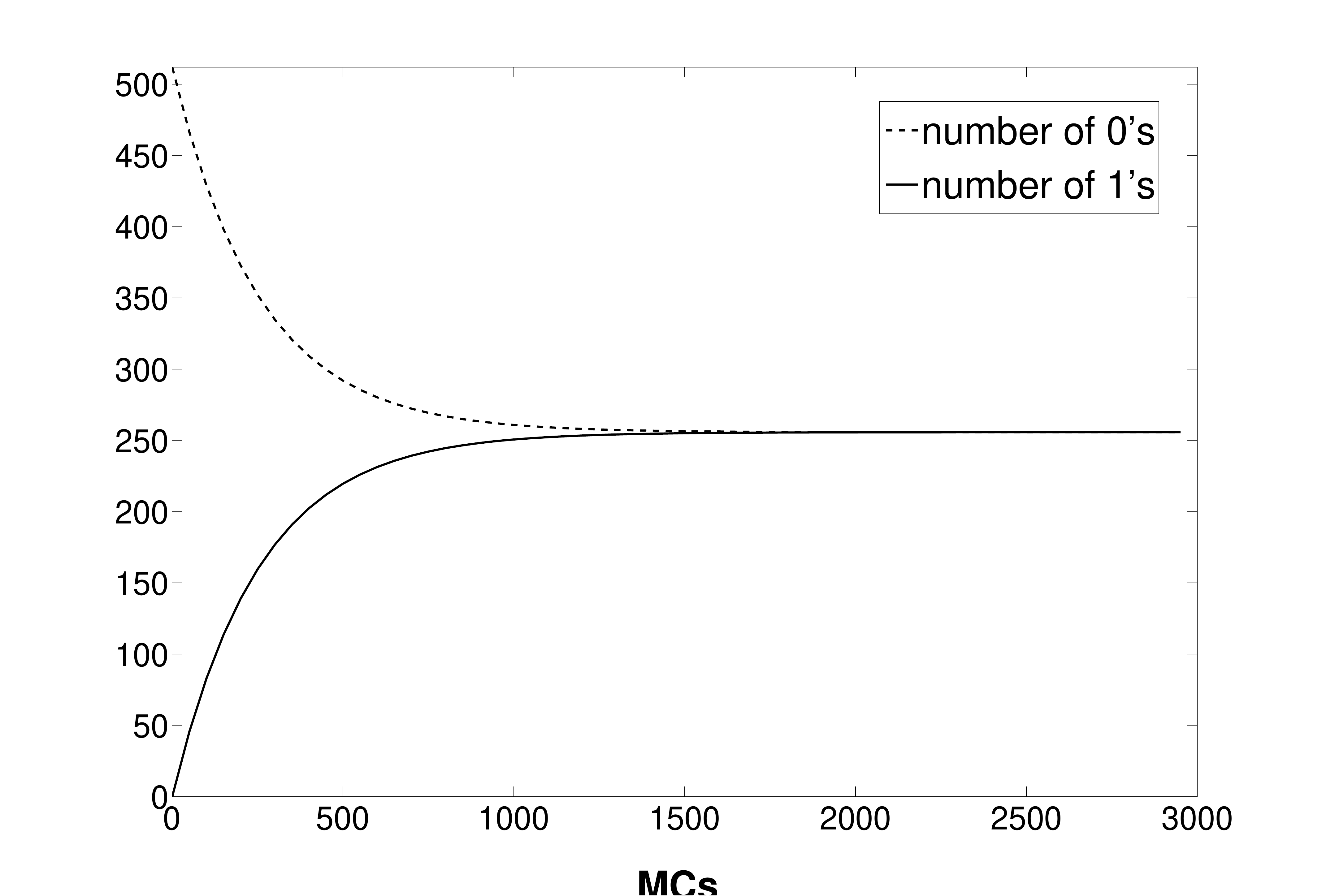}
\includegraphics[angle=0,width=0.494\textwidth]{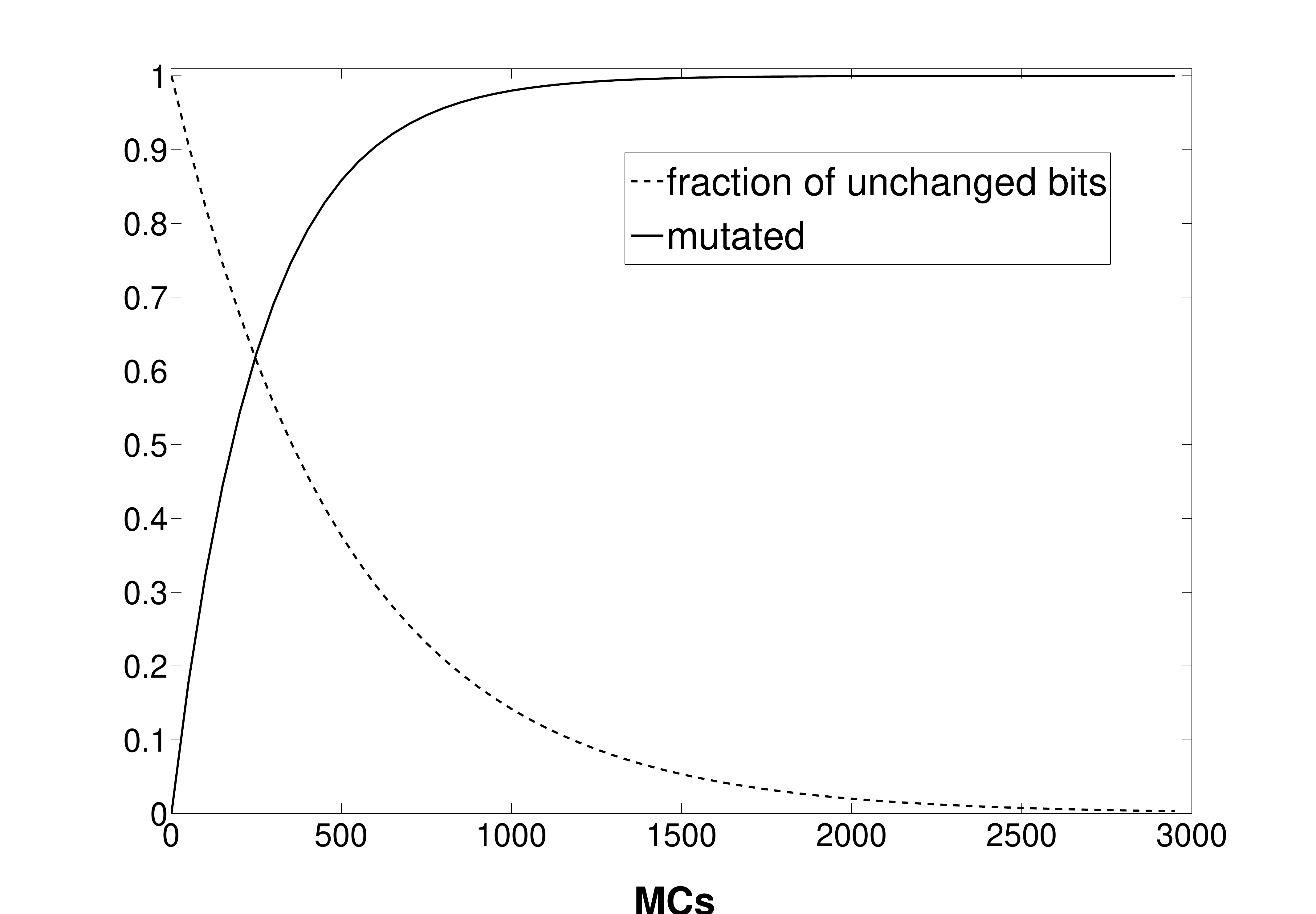}
\caption{The distribution of the requirements of the randomly changing environment in time. Simulations starts with all functions required by the environment. One function is randomly chosen at each Monte Carlo step and it flops from 0 to 1 or vice versa. After about 1500 MCs the number of neutral functions equals the number of required functions (left plot). In the right plot, the y-axis presents the fraction of bits in the environment pattern which were not changed during a given period of simulation, measured in MCs (x-axis) and the fraction of positions in the genome which were hit by the mutation during the same period of time.}
\label{f5}
\end{figure}
To check the effect of the environmental changes, a defined part of loci in the genomes has been declared as neutral at the start of simulations (the first version of the environment changes). After a given period of time, the environment requires the functions of those loci again while some other loci become neutral, and after the second period, the functions of those loci are again required while the last group become neutral. In fact, in this version, the environmental changes are cyclic. Short periods of neutrality have not a deleterious effect because all individuals are checked during their life for the functionality of all genes. But even in such a case, some negative effects of environmental changes could be observed. It is connected with some delay of the phenotypic expression of the genetic defects. Defective genomes are eliminated later than normally exploiting the environment in vain or even transferring the defect to the next generation. If the environment does not require a given function for longer periods - individuals can accumulate the defective alleles which kill their offspring when environment is again demanding the function. The fluctuations of the population sizes caused by these changes are shown in Fig. \ref{f4}. What is highly unintuitive, the probability of population survival is higher when the cross over rate is lower in case of the cyclic environmental changes. Close to the critical values of parameters, when populations are close to the extinction, the surviving individuals are not dispersed evenly on the whole territory. They are grouped. The snapshots just before and after the changes are shown in Fig. \ref{f3}. For more detailed visualizations and animations please visit: http://www.smorfland.uni.wroc.pl/sympatry/env. All individuals in a given group possess functional genes actually required by the environment, which suggest that they belong to one family. To prove that hypothesis, we have checked the genetic relations back to the fifth generation. All other squares occupied by individuals related up to the fifth generation will highlight. Since the density of populations drops and, additionally, organisms forming groups are genetically related, the speciation is accelerated.  
\begin{figure}
\includegraphics[angle=0,width=\textwidth]{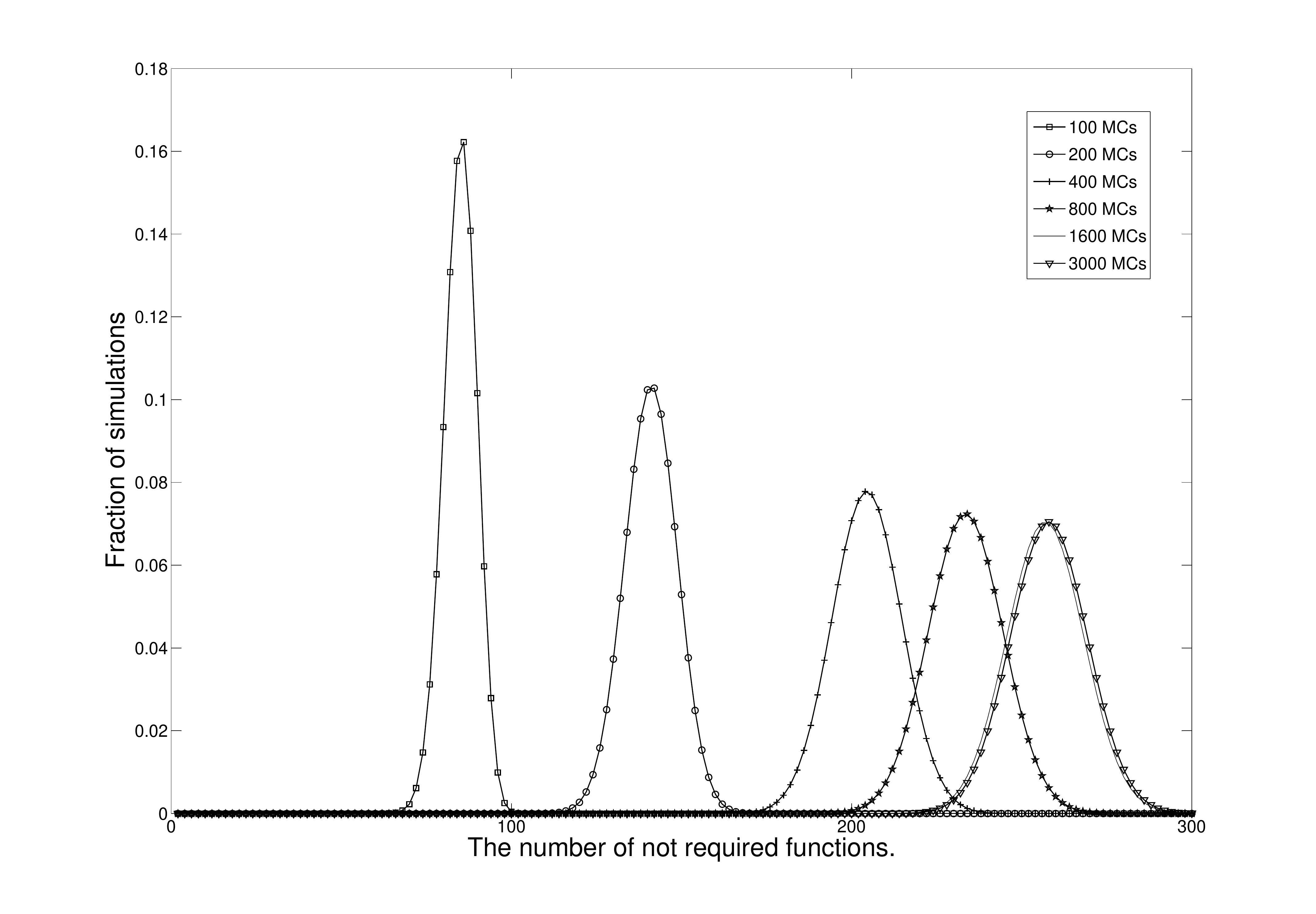}
\caption{The expected distribution of the function requirement after different time of simulations (in MCs). Simulations start with all functions required.}
\label{f6}
\end{figure}
\begin{figure}
\includegraphics[angle=0,width=\textwidth]{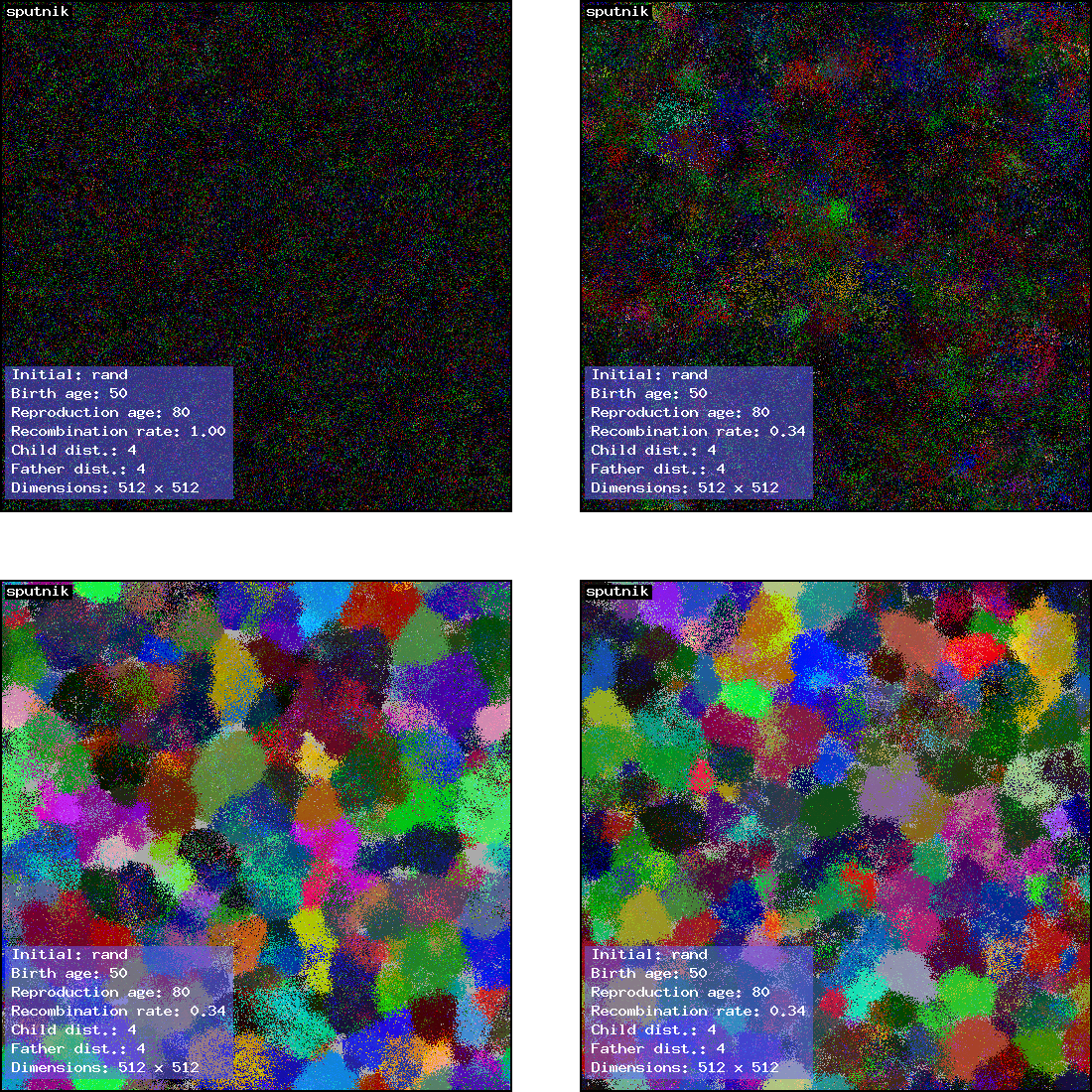}
\caption{The snapshots of populations evolved under different regimes. Upper panels in the constant environment (left lattice with cross over rate 1.0 per chromosome pair, right lattice with cross over rate 0.34). Lower panel with the same crossover rate (left lattice with randomly changing environment - 2 bits each MCs, right lattice in periodically changing environment (see text for more details).}
\label{f7}
\end{figure}

Random changes of the environment were introduced by declaring that the constant number of randomly chosen bits in the environment pattern is changed every MCs and the probability of changes in both directions are equal. The dynamic of such changes is shown in Fig. \ref{f5}. In particular, when only one bit is changed every MCs, after about 1500 MCs the environmental requirements is already close to the equilibrium - almost half of functions are neutral (bits in the environment pattern set to 1) and half are required (bits in the environment pattern set to 0). The expected distribution of  a ,,bit survival'', which means what is the fraction of bits of the original environmental pattern which have not been changed, drops exponentially with the number of MCs as it is shown in the Fig. \ref{f5} (the fraction of unchanged bits). Some functions stay neutral for a very long time (statistically, 2\% of functions can stay neutral for more than 2000 MCs). Since mutation rate is of the order of 2 mutations per haplotype per generation, the large fraction of genes coding for functions which are neutral for very long time accumulates mutations and kills the carriers after switching on the requirement for the very function. The process of switching on and off of the environmental requirements is stochastic and the expected distributions of the environmental requirements after different time of simulations are shown in Fig. \ref{f6}. 
The effect of the environmental changes on speciation is shown in Fig. \ref{f7}. The upper two panels show the population evolution in the constant environment. The only difference in parameters between the two populations is in the recombination frequency. In the left one, there is C=1.0 while in the right one C=0.34. In the lower panels both populations evolved under the same parameters, but the right one was in the periodically changing environment while the left one evolved in the randomly changing environment with requirements for two functions changing every MCs. Comparing the upper panel with the lower one, it could be concluded that the environmental changes accelerate the sympatric speciation.

\section{Acknowledgements}

We thank D. Stauffer for comments and discussions. The work has been done in the frame of European programs: COST Action MP0801, FP6 NEST - GIACS and UNESCO Chair of Interdisciplinary Studies, University of Wroc\l{}aw.

Calculations have been carried out in Wroc\l{}aw Centre for Networking and Supercomputing (http://www.wcss.wroc.pl), grant \#102.

\end{document}